\documentclass[journal]{IEEEtran}
\usepackage{amsmath,bm}

\usepackage{cite}

\ifCLASSINFOpdf
  \usepackage[pdftex]{graphicx}
\else
\fi
\usepackage{amsmath}
\usepackage{amssymb}
\usepackage{array}

\usepackage[caption=false,font=footnotesize]{subfig}
\usepackage{multirow}

\usepackage{threeparttable}
\usepackage{stfloats}

\begin{document}
%
\title{Multi-Rate Nyquist-SCM for C-Band 100Gbit/s Signal over 50km Dispersion-Uncompensated Link}
%
%
%

\author{Haide~Wang,
	Ji~Zhou,
	Jinlong Wei,
	Dong~Guo,
	Yuanhua~Feng,
	Weiping~Liu,
	Changyuan~Yu,
	Dawei~Wang,	
	and~Zhaohui~Li
	
\thanks{Manuscript received; revised. This work was supported in part by National Key R\&D Program of China (2019YFB1803500); National Natural Science Foundation of China (62005102, U2001601); Natural Science Foundation of Guangdong Province (2019A1515011059); Guangzhou Basic and Applied Basic Research Foundation (202102020996); Fundamental Research Funds for the Central Universities (21619309); Open Fund of IPOC (BUPT) (IPOC2019A001). (Corresponding authors: Ji Zhou)}
	
\thanks{H. Wang, J. Zhou, Y. Feng and W. Liu are with Department of Electronic Engineering, College of Information Science and Technology, Jinan University, Guangzhou 510632, China (e-mail: 1834041007@stu2018.jnu.edu.cn; zhouji@jnu.edu.cn; favinfeng@163.com; wpl@jnu.edu.cn).}
	
\thanks{J. Wei is with Huawei Technologies Duesseldorf GmbH, European Research Center, Germany (e-mail:jinlongwei2@huawei.com).}	

\thanks{D. Guo is with School of Information and Electronics, Beijing Institute of Technology, Beijing 100081, China (e-mail: 7520190141@bit.edu.cn).}
	
\thanks{C. Yu is with the Department of Electronic and Information Engineering, The Hong Kong Polytechnic University, Hong Kong (e-mail: changyuan.yu@polyu.edu.hk).}
	
\thanks{D. Wang and Z. Li are with the Guangdong Provincial Key Labratory of Optoelectronic Information Processing Chips and Systems, Sun Yat-sen University, Guangzhou 510275, China. Also with the State Key Laboratory of Optoelectronic Materials and Technologies. And also with the Southern Marine Science and Engineering Guangdong Laboratory (Zhuhai), 519000, China (e-mail: wangdw9@mail.sysu.edu.cn, lzhh88@mail.sysu.edu.cn).}}
%
%

\markboth{}%
{JOURNAL OF LIGHTWAVE TECHNOLOGY}
%



\maketitle

\begin{abstract}
In this paper, to the best of our knowledge, we propose the first multi-rate Nyquist-subcarriers modulation (SCM) for C-band 100Gbit/s signal transmission over 50km dispersion-uncompensated link. Chromatic dispersion (CD) introduces severe spectral nulls on optical double-sideband signal, which greatly degrades the performance of intensity-modulation and direct-detection systems. Based on the prior knowledge of the dispersive channel, Nyquist-SCM with multi-rate subcarriers is proposed to keep away from the CD-caused spectral nulls flexibly. Signal on each subcarrier can be individually recovered by a digital signal processing, including the feed-forward equalizer with no more than 31 taps, a two-tap post filter, and maximum likelihood sequence estimation with one memory length. Combining with entropy loading based on probabilistic constellation shaping to maximize the capacity-reach, the C-band 100Gbit/s multi-rate Nyquist-SCM signal over 50km dispersion-uncompensated link can achieve 7\% hard-decision forward error correction limit and average normalized generalized mutual information of 0.967 at received optical power of $-4$dBm and optical signal-to-noise ratio of 47.67dB. In conclusion, the multi-rate Nyquist-SCM shows great potentials in solving the CD-caused spectral distortions.
\end{abstract}


%
\IEEEpeerreviewmaketitle

\section{Introduction}
%
%

\IEEEPARstart{I}{n} recent years, network applications such as 4K/8K high-definition television, augmented reality/virtual reality, and edge cloud computing are driving the growth of the global data traffics \cite{zhou2019beyond}. To satisfy the ever-increasing capacity demands, high-speed intensity-modulation and direct-detection (IM/DD) optical systems have been paid much attentions, which keep the advantages of low cost, low power consumption and small footprint \cite{zhong2018digital, cheng2018recent, liu2018dd}. However, compared to the O-band IM/DD optical systems, the systems operating at C-band with relative low link loss are facing with much more severe power fading challenge caused by chromatic dispersion (CD) \cite{chagnon2019direct, wei2018challenges, hong2021numerical}.

In order to eliminate distortions caused by CD, dispersion-compensation fiber/module (DCF/DCM) can be used \cite{eiselt2016evaluation, zhang2017eml}. However, DCM not only complicates the link configuration, but also increases link loss. In addition, other solutions have drawn much attention to resisting CD, which mainly include optical single-sideband (SSB) modulation \cite{lee2016112,zhang2014c, zhang2015c, wan201764}, Kramers–Kronig (KK) receiver \cite{li2018spectrally, le2019experimental,le20195, mecozzi2016kramers} and CD pre-compensation \cite{zhang2016transmission, erkilincc2015spectrally, torres2020100+}. However, these schemes require either more complicated transmitters or receivers compared to IM/DD systems. For optical double-sideband (DSB) signal, the advanced digital signal processing (DSP) can effectively deal with CD. However, high-complexity DSP algorithms are required to resist CD-caused spectral nulls, including the decision feedback equalizer \cite{tang2019c, xin202050}, Tomlinson-Harashima pre-coding \cite{rath2017tomlinson}, and maximum likelihood sequence estimation (MLSE) \cite{zhou2020c, zhou2021processing}. Discrete multi-tone (DMT) with adaptive bit and power loading (ABPL) has a CD resistance but its peak-to-average power ratio (PAPR) is relatively high \cite{nadal2014dmt, le2020100gbps}. Different solutions to CD in the direct-detection systems are summarized in Table \ref{Solutions}. 

\begin{table*}[!t]
	\centering
	\begin{threeparttable}[b]
	\renewcommand\arraystretch{1.2}
	\caption{Different solutions to CD in the direct-detection systems.}
	\begin{tabular}{cccc}
		\hline 
		\hline
	    References & Solutions  & Required Components & Required DSP \\ 
		\hline
		\cite{eiselt2016evaluation, zhang2017eml} &  Optical compensation & DCF or DCM  &  FFE, DFE, Look-up table \\
		
		\begin{tabular}[c]{@{}c@{}}\cite{lee2016112, zhang2014c}\\ \cite{wan201764, zhang2015c}\end{tabular} & \begin{tabular}[c]{@{}c@{}}{Optical SSB}\end{tabular}  & \begin{tabular}[c]{@{}c@{}}Optical filter\\  DD-MZM or I/Q-MZM, Two DACs\end{tabular}
		& \begin{tabular}[c]{@{}c@{}} NLE \\ Hilbert transform, NLE\end{tabular} \\

        \begin{tabular}[c]{@{}c@{}}\cite{li2018spectrally, le2019experimental} \\ \cite{le20195, mecozzi2016kramers}\end{tabular} 
        & KK \& Optical SSB   &  \begin{tabular}[c]{@{}c@{}} DD-MZM or I/Q-MZM, Two DACs, High-sampling ADC \\ Tone laser, High-sampling ADC
        \end{tabular} & \begin{tabular}[c]{@{}c@{}} KK receiver, CD compensation, NLE
        \end{tabular}\\ 
        
		\cite{zhang2016transmission, erkilincc2015spectrally, torres2020100+} & CD pre-compensation & DD-MZM or I/Q-MZM, Two DACs &  CD compensation,  FFE, MLSE \\
		
		\cite{tang2019c, xin202050, rath2017tomlinson, zhou2020c, zhou2021processing}& IM (OOK, PAM)    & DML or MZM, One DAC &  FFE, DFE, MLSE, THP \\ 
		\cite{nadal2014dmt, le2020100gbps} & IM (DMT) & DML or MZM, One DAC &  ABPL, FFT  \\ 
		This work & IM (Multi-rate Nyquist-SCM) & DD-MZM @ PPM, One DAC &  FFE, MLSE \\ 
		\hline \hline 
	\end{tabular}\label{Solutions}
	\begin{tablenotes}
       \footnotesize
       \item[] DD-MZM: dual-drive Mach–Zehnder modulator. I/Q-MZM: in-phase/quadrature Mach–Zehnder modulator. DAC: digital-to-analog converter. ADC: analog-to-digital converter. NLE: nonlinear equalizer. DML: directly modulated laser. FFT: fast Fourier transform. PPM: push-pull mode.
     \end{tablenotes}
     \end{threeparttable}
\end{table*}

Recently, optical SSB Nyquist-subcarriers modulation (SCM) has attracted much attention because of the strong CD resistance and high spectral efficiency. The PAPR of Nyquist-SCM with few subcarriers is also lower than that of DMT with much more subcarriers guided by the central limit theorem \cite{park2000papr}. Literature on Nyquist-SCM is reviewed before a novel Nyquist-SCM is proposed. A 112Gbit/s IM/DD Nyquist-SCM over 4km SSMF transmission is demonstrated in \cite{cartledge2014100}. $6\times25$Gbit/s and $7\times28$Gbit/s wavelength-division multiplexing SSB Nyquist-SCM transmissions are presented in \cite{erkilincc2016spectrally, erkilinc2015performance}. In \cite{bo2018toward}, the performance of Nyquist-SCM using upsampling-free KK algorithm is also evaluated. Nyquist-SCM with several subcarriers has strong tolerance to channel distortions \cite{vassilieva2019enabling}. In addition, Nyquist-SCM has been used in long-haul coherent optical transmission systems \cite{qiu2014subcarrier}. The commercial 800G application specific integrated circuits have  also adopted SCM technology \cite{sun2020800g}. Literature review shows that Nyquist-SCM is a promising technique with strong robustness against channel distortions and high spectral efficiency. 

Compared with optical SSB signals, optical DSB signal is easier to be generated and has a higher optical signal-to-noise ratio (OSNR), although SSB signal is robust against CD-caused power fading. In this paper, we propose an optical DSB multi-rate Nyquist-SCM to keeps away from the CD-caused spectral nulls flexibly. Therefore, signal on each subcarrier can be recovered individually by adaptive channel-matched detection (ACMD) algorithm \cite{wang2020adaptive}, including the feed-forward equalizer (FFE) with no more than 31 taps, a two-tap post filter (PF) and MLSE with one memory length. Moreover, by combining entropy loading based on probabilistic constellation shaping (PCS) with SCM, it can maximize the capacity-reach \cite{sun2020800g}. The main contributions of this paper are as follows:
\begin{itemize}
\item The multi-rate Nyquist-SCM system is proposed, which successfully combines the advanced technologies of Nyquist-SCM, multi-rate subcarriers, ACMD algorithm and the entropy loading based on PCS.

\item To the best of our knowledge, we present the first C-band 100Gbit/s IM/DD optical multi-rate Nyquist-SCM transmission over a 50km dispersion-uncompensated link achieving 7\% hard-decision forward error correction (HD-FEC) limit and average normalized generalized mutual information (NGMI) of 0.967.
\end{itemize}

\begin{figure}[!t]
	\centering
	\includegraphics[width = 3.0 in]{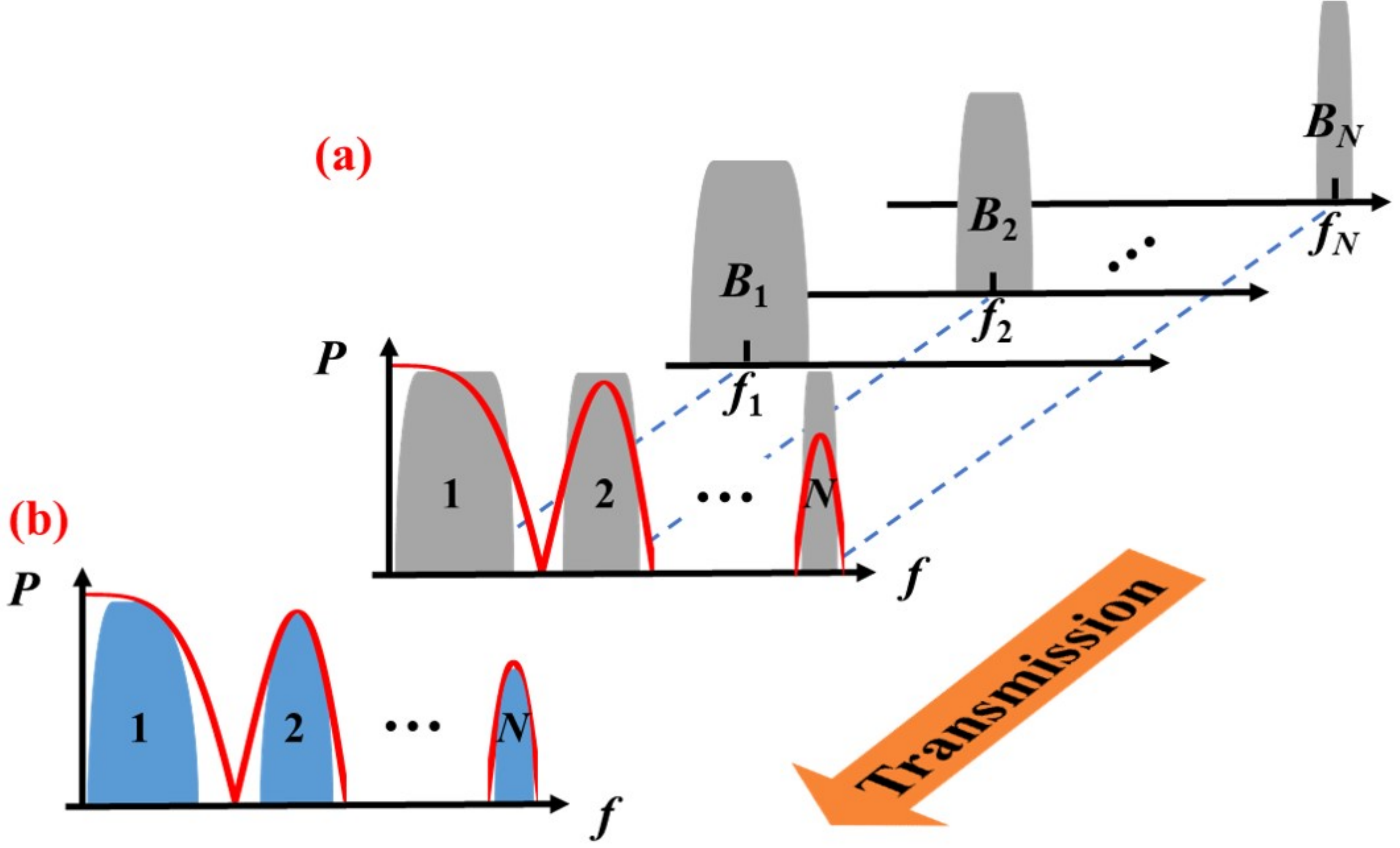}
	\caption{The schematic diagram of multi-rate Nyquist-SCM signal spectrum at (a) the transmitter side and (b) the receiver side, respectively.}
	\label{SP}
\end{figure}

The remainder of the paper is organized as follows. In Section \ref{Principle}, DSP of the multi-rate Nyquist-SCM is given. In Section \ref{ES}, the experimental setups of C-band 100Gbit/s IM/DD optical multi-rate Nyquist-SCM system over 50km dispersion-uncompensated link are presented. The experimental results and discussions are demonstrated in Section \ref{ER}. Finally, the paper is concluded in Section \ref{Conclusion}.

\section{DSP of multi-rate Nyquist-SCM}\label{Principle}

After the square-law detection, the frequency response of fiber dispersive channel can be expressed as \cite{zhou2016transmission}
\begin{equation}
H(f)=\cos \left(2 \pi^{2} \beta_{2} L f^{2}\right)
\label{eq.1}
\end{equation}
where $\beta_{2}$ is group velocity coefficient and $L$ is the fiber length. This CD-caused power fading will result in spectral nulls in signal spectrum \cite{zhou2021burst}. Fig. \ref{SP} shows the schematic diagram of multi-rate Nyquist-SCM signal spectrum at (a) the transmitter side and (b) the receiver side, respectively. Due to limited bandwidth of devices and the CD of the fiber link, the frequency response of the channel is a cosine function with an overall downward trend as the red line shown in Fig. \ref{SP}. 

Based on the dispersive channel response, the spectral nulls are searched automatically to determine the band count $N$ and the range of each band. Baud rate of the signal on each subcarrier is the difference between two frequency points, which attenuate about 10dB from the point with maximum response within the band. If a more effective but complicated DSP algorithm is used, the points with greater attenuation can be used. Then central frequency of each band is set to the middle of the two frequency points. After Nyquist shaping and frequency up-conversion, the central frequency of $i^{\rm{th}}$ band signal is $f_i$ and the bandwidth is $B_i$, where $i = 1, 2,..., N$. The proposed multi-rate Nyquist-SCM system can flexibly adjust band count, bandwidth and central frequency of each band to keep away from CD-caused spectral nulls. After transmission, each band signal does not suffer from CD-caused spectral nulls. Therefore, the proposed multi-rate Nyquist-SCM system naturally has a strong resistance to CD. 

\begin{figure*}[!t]
	\centering
	\includegraphics[width = 0.9\linewidth]{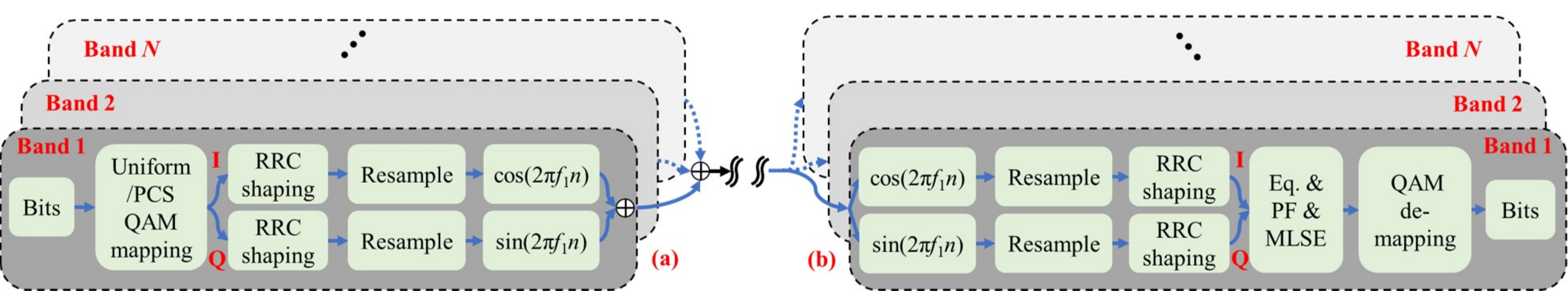}
	\caption{The block diagram of DSP for the $N$-band multi-rate Nyquist-SCM system at (a) the transmitter side and (b) the receiver side.}
	\label{DSP}
\end{figure*}

Figure \ref{DSP} shows DSP of the $N$-band multi-rate Nyquist-SCM system at (a) the transmitter side and (b) the receiver side. $N$ tributaries bits are mapped to the uniform quadrature amplitude modulation (QAM) with different modulation order or PCS QAM symbols with different entropy to make full use of the SNR. The uniform modulation format of each band is optimized according to the SNR of each band. The mutual information of uniform QAM is used as PCS QAM initial entropy $H_{0,i}$ for the $i^{\rm{th}}$ band. In order to reach the target rate $RS_{\rm{target}}$, the entropy is modified as
\begin{equation}
H_{i}=H_{0, i}+\left(RS_{\rm{target}}-\sum_{t=1}^{N} H_{0, t} \times R S_{t}\right) / \sum_{t=1}^{N} R S_{t}.
\end{equation}
Then the corresponding shaping factor for PCS QAM is chosen by the look-up table. For the $i^{\rm{th}}$ band, the in-phase (I) and quadrature (Q) components of QAM symbol are up-sampled by a factor of $\left\lfloor f_{s} / R S_{i}\right\rfloor$ and filtered by square-root raised cosine (RRC) filters to realize up-sampling and Nyquist-shaping. $f_{s}$ is the sampling rate of digital-to-analog converter (DAC) and $\lfloor\cdot\rfloor$ is the round floor operation. The signal is resampled by a factor of $f_{s} /\left(R S_{i} \times\left\lfloor f_{s} / R S_{i}\right\rfloor\right)$. Then the baseband signal is shifted to the central frequency of $f_i$. Finally, signals of all bands are combined and reconstructed to the transmitted analog signal. 

\begin{figure*}[!t]
	\centering
	\includegraphics[width = 0.9\linewidth]{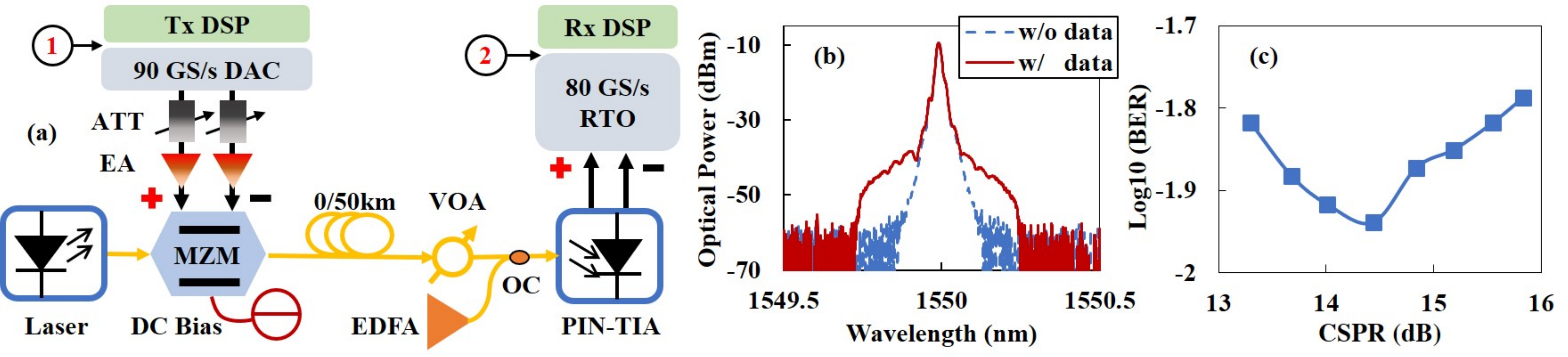}
	\caption{(a) The experimental setups of C-band 100Gbit/s IM/DD optical multi-rate Nyquist-SCM system transmission over dispersion-uncompensated links. (b) Optical spectrum of the received signal over 50km SSMF transmission and (c) BER versus CSPR of multi-rate Nyquist-SCM signal at ROP of $-6$dBm.}
	\label{EX}
\end{figure*}

At the receiver side, the electrical intermediate frequency signal at each band is shifted to the baseband. After being resampled, the baseband signals are filtered by a matched filter. After down sampling, the received I and Q components make up the $M$-QAM symbol for equalization. The signal on each subcarrier can be recovered by the ACMD algorithm of single-carrier modulation, which is proposed in our previous work \cite{wang2020adaptive}. First, FFE eliminates the linear distortions and cycle slip issue \cite{wei2017multi}, tap coefficients of which are trained by least mean square (LMS) algorithm and tracked by decision directed-LMS algorithm. Because the proposed multi-rate Nyquist-SCM signal has the advantage of strong resistance to CD, DFE is not required to compensate the spectral nulls. Then a followed two-tap PF can whiten the colored noise after equalization. The output of the $i^{\rm{th}}$ band PF is 
\begin{equation}
p_i(n) = q_i(n) + \alpha \times q_i(n-1)
\end{equation}
where $\alpha$ is the tap coefficient of PF. The known inter-symbol interference (i.e., $\alpha \times q_i(n-1)$) can be eliminated by MLSE with one memory length. Finally, the QAM symbol is de-mapped to bits.

\section{Experimental Setups}\label{ES}

The experimental setups of C-band 100Gbit/s IM/DD optical multi-rate Nyquist-SCM system over dispersion-uncompensated links is shown in Fig. \ref{EX}(a). First of all, a 64Gbit/s on-off keying (OOK) preamble was launched. Feedback of the OOK preamble from the receiver to the transmitter was used to estimate the channel frequency response by using fast Fourier transform at the initial stage. Then the counts and positions of the spectral nulls within the bandwidth of interest were used to design the multi-rate Nyquist-SCM signal according to the prior knowledge of channel. 

The band count of the multi-rate Nyquist-SCM system transmission over 50km standard single-mode fiber (SSMF) was 7. There were 378260 bits divided into 7 tributaries for generating the multi-rate Nyquist-SCM, which was designed based on the estimated channel response. And the rate $RS_i$ was set to 7.01, 5.11, 3.81, 2.76, 2.9, 2.62 and 1.92GBaud, respectively. The central frequency $f_i$ of each band was set to 3.9, 12, 17.3, 21.3, 24.5, 27.4 and 29.95GHz, respectively. As shown in Equation \ref{eq.1}, the higher the frequency of spectral null is, the faster the channel response nearby changes. Therefore, the roll-off factor of RRC filter was set to 0.1 for the first three bands and 0.01 for the rest four bands, respectively.

After DSP at the transmitter side as shown in Fig. \ref{DSP}(a), the frame was uploaded into a DAC with resolution of 8 bit, 90GSa/s sampling rate and 16GHz 3dB bandwidth. After being amplified by the electrical amplifiers (EA, Centellax OA4SMM4) followed the 6-dB attenuators (ATT), a 40Gbps Mach-Zehnder modulator (MZM, Fujitsu FTM7937EZ) @ push-pull mode with two differential inputs was used to modulate the amplified multi-rate Nyquist-SCM signal on a continuous wave optical carrier at 1550.02nm for generating optical DSB multi-rate Nyquist-SCM signal. The generated optical multi-rate Nyquist-SCM signal was fed into the 0/50km SSMF, respectively. The launch optical power was set to 6.56dBm (i.e., the maximum launch optical power of the device). The maximum received optical power (ROP) was approximately $-4$dBm and total link loss was approximately 10.66dB over 50km SSMF transmission. 

\begin{figure}[!t]
	\centering
	\includegraphics[width =0.847\linewidth]{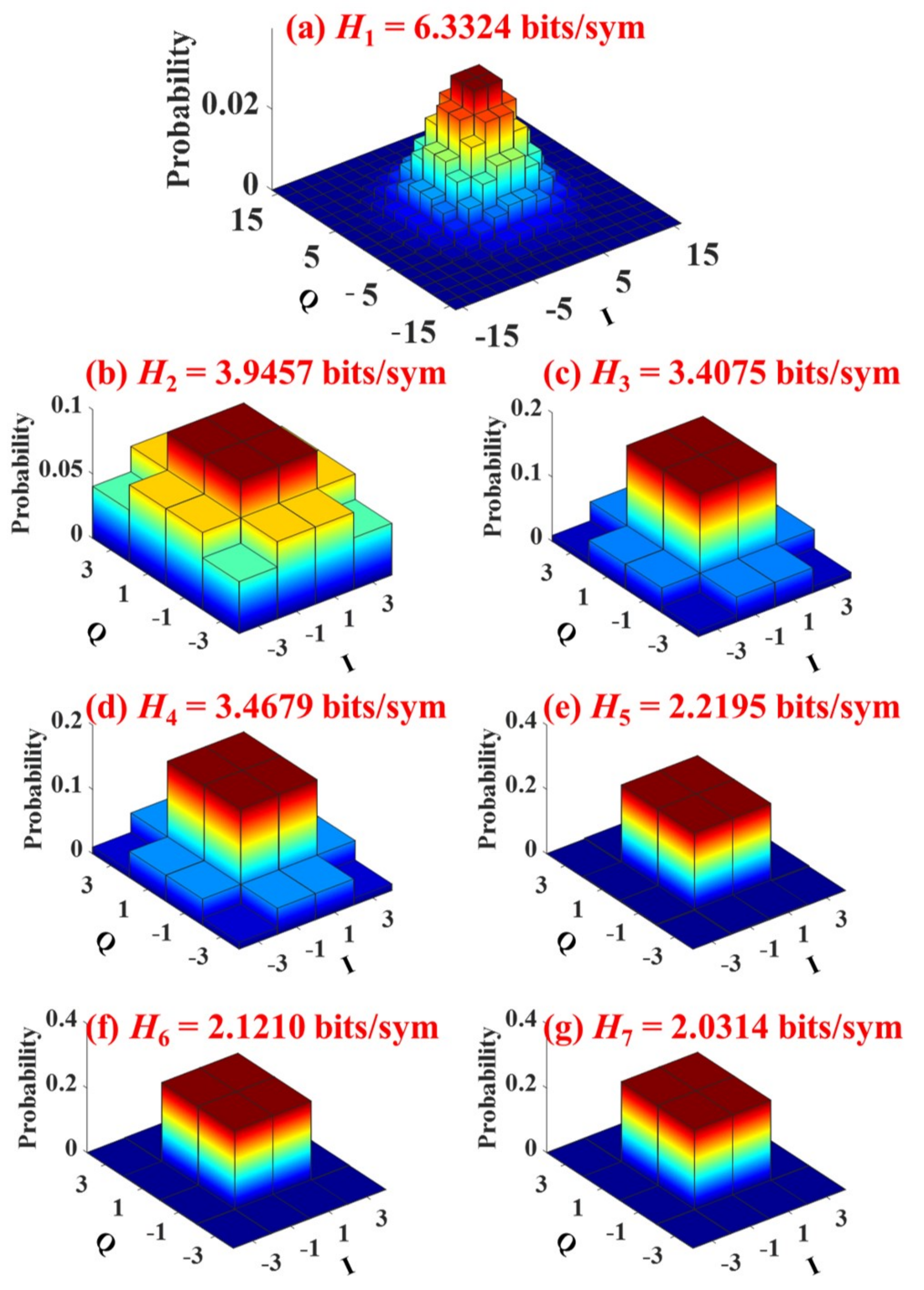}
	\caption{The probability of the constellation points of the PCS QAM adopted by each band.}
	\label{Probability}
\end{figure}

\begin{figure}[!t]
	\centering
	\includegraphics[width = 2.81 in]{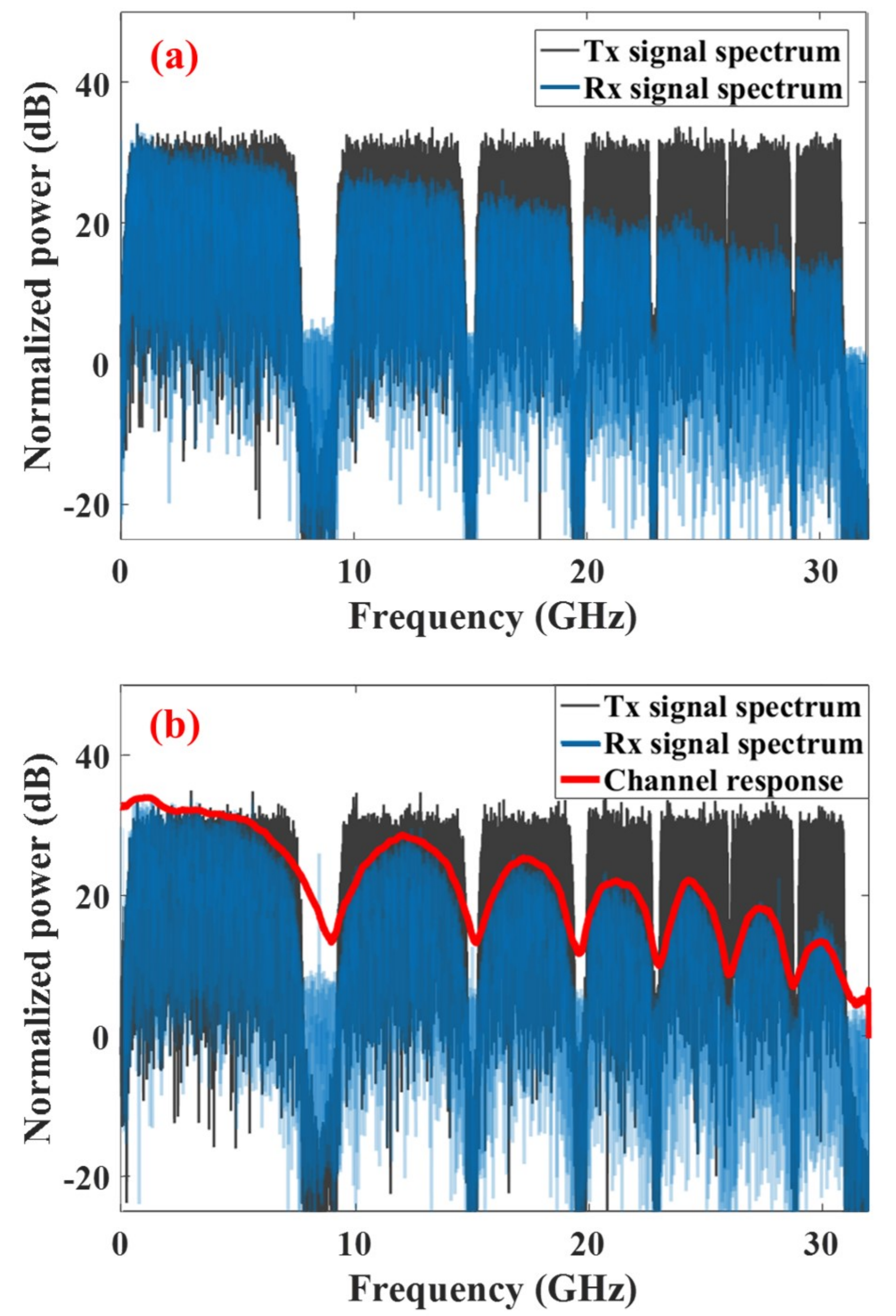}
	\caption{Electrical frequency spectrum of the transmitted signal and received signal of the 100Gbit/s multi-rate Nyquist-SCM system over (a) OBTB transmission and (b) 50km SSMF transmission and estimated channel response.}
	\label{Tx_signal}
\end{figure}

At the receiver side, a variable optical attenuator (VOA) was used to adjust the ROP and an erbium-doped fiber amplifier (EDFA) was used as the amplified spontaneous emission (ASE) noise source to adjust the OSNR via an optical coupler (OC). Then optical multi-rate Nyquist-SCM signal was converted into an electrical signal by a 31GHz P-type-intrinsic-N-type diode with trans impedance amplifier (PIN-TIA, Finisar MPRV1331A), which has a differential output. The differential output signals were fed into a 80GSa/s real-time oscilloscope (RTO) with cut-off bandwidth of 36GHz to implement A/D conversion. Then the digital multi-rate Nyquist-SCM signal was decoded by the off-line processing, including resampling, synchronization, multi-rate Nyquist-SCM signal decoding as shown in Fig. \ref{DSP}(b) and bit error rate (BER) calculation.

The optical spectrum of the received signal over 50km SSMF transmission at the final ROP of about $-4$dBm is shown in Fig. \ref{EX}(b). The OSNR (0.1nm noise bandwidth) is about 47.67dB, which contains the power of the optical carrier. Therefore, the carrier-to-signal power ratio (CSPR) optimization is important for the system performance. A large CSPR can suppress signal-signal beat interference, while it also reduces the receiver sensitivity. CSPR is adjusted by changing the direct current (DC) bias voltage, while the amplitude of driving signal keeps unchanged. Fig. \ref{EX}(c) shows BER versus CSPR of the multi-rate Nyquist-SCM signal. There will be different maximum ROPs when different bias voltages are applied. Hence ROP is fixed at $-6$dBm. The optimal CSPR is about 14.44dB at the 2.3V bias voltage. 

\section{Experimental Results And Discussion}
\label{ER}

\begin{figure*}[!t]
	\centering
	\includegraphics[width = 0.98\linewidth]{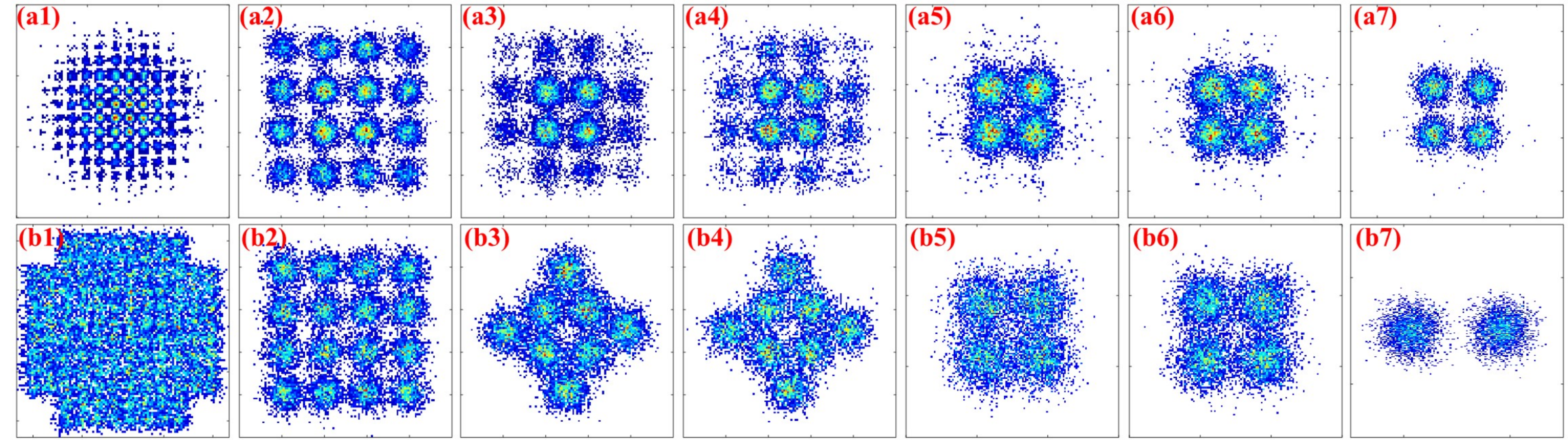}
	\caption{Constellations of the equalized signal over 50km SSMF transmission at ROP of approximately $-4$dBm adopting (a1)-(a7) the PCS QAM and (b1)-(b7) the uniform QAM of each band.}
	\label{Constellation}
\end{figure*}

Due to the limited bandwidth, capacity of bands in high frequency are lower than those close to the low frequency. The uniform QAM with different order and PCS QAM with different entropy are used respectively. Then 128QAM, 16QAM, 8QAM, 8QAM, 4QAM, 4QAM and binary phase shift keying is used for the band from $1^{\rm{st}}$ to $7^{\rm{th}}$, respectively. The link rate is 102.18Gbit/s. Based on the PCS with constant composition distribution matching, the QAM formats with various probabilistic distributions for entropy loading are allocated to each band. PCS 256QAM is used for $1^{\rm{st}}$ band and PCS 16QAM with different entropy are used for the rest bands. Figs. \ref{Probability}(a)-(g) show the probability of the constellation points of the PCS QAM adopted by each band, respectively. The entropy of each band is also attached and the target rate is 103Gbit/s, which includes the overhead of training sequence. The lengths of training sequence for these 7 bands are set to 500, 300, 300, 300, 200, 200 and 100, respectively. After deducting training overhead, the data rates are about 100.377Gbit/s and 100.301Gbit/s for PCS QAM and uniform QAM, respectively.

Figure \ref{Tx_signal} shows the electrical frequency spectrum of transmitted (Tx) signal and received (Rx) signal of the 100Gbit/s multi-rate Nyquist-SCM system over (a) optical back-to-back (OBTB) and (b) 50km SSMF transmission and the estimated dispersive channel response. Tx signal spectrum corresponds to spectrum of the multi-rate Nyquist-SCM signal at Point $1$ in Fig.\ref{EX} and Rx signal spectrum corresponds to that at Point $2$. Total bandwidth of the multi-rate Nyquist-SCM signal is approximately 31GHz. As the Fig. \ref{Tx_signal}(a) shows, only the limited bandwidth causes the power fading of the received signal over OBTB transmission. However, after 50km SSMF transmission, the signal suffers from power fading caused by both limited bandwidth and CD as shown in Fig. \ref{Tx_signal}(b). 

Constellations of the equalized signal over 50km SSMF transmission at ROP of about $-4$dBm adopting the PCS QAM and uniform QAM of each band are shown in Fig. \ref{Constellation}. As shown in Figs. \ref{Constellation}(a1)-(a7), the occurrence probabilities of the PCS QAM outer constellation points are lower than those of the inner points, while those of the uniform QAM are approximately the same as shown in Figs. \ref{Constellation}(b1)-(b7). The 100Gbit/s multi-rate Nyquist-SCM signal is well-designed based on the estimated channel response to keep away from CD-caused spectral nulls. Therefore, the multi-rate Nyquist-SCM has a strong resistance to CD and a FFE with no more than 31 taps can equalize the signal on each subcarrier. The tap numbers of FFEs are 31 for the first two bands and 21 for the rest five bands, respectively. The step sizes are $5\times10^{-3}$ and $10^{-3}$ for LMS algorithm and DD-LMS algorithm, respectively. According to the filtering of each band, the PF coefficient $\alpha$ is 0.3, 0.4, 0.4, 0.4, 0.5, 0.5 and 0.2, respectively.

\begin{figure}[!t]
	\centering
	\includegraphics[width = 0.9\linewidth]{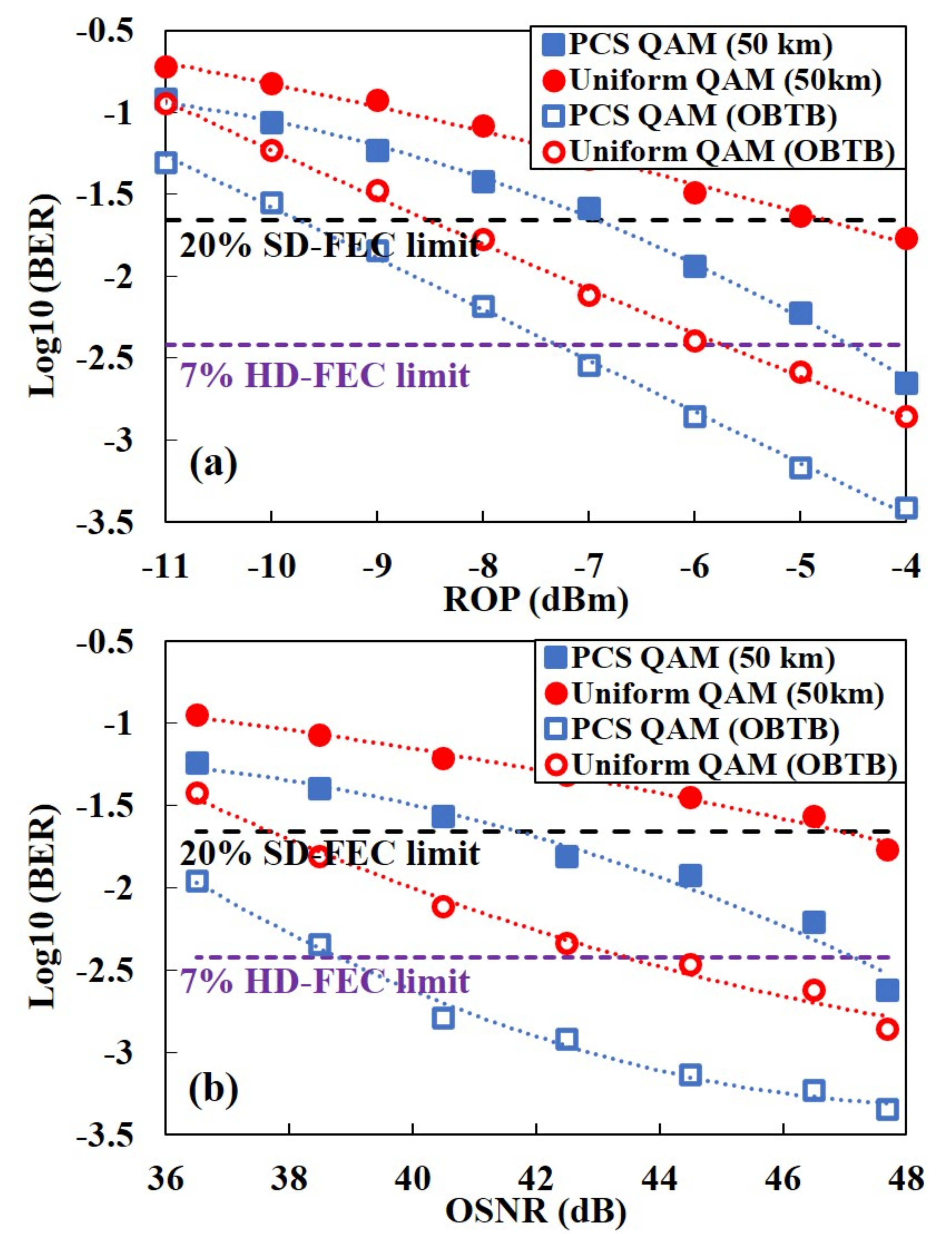}
	\caption{BER performance versus (a) ROPs and (b) OSNRs for 100Gbit/s IM/DD optical multi-rate Nyquist-SCM system adopting uniform QAM and PCS QAM schemes over OBTB and 50km SSMF transmission, respectively.}
	\label{BER}
\end{figure}

BER performance versus ROPs for 100Gbit/s IM/DD optical multi-rate Nyquist-SCM system is shown in Fig. \ref{BER}(a). After OBTB transmission, BER is below the 7\% HD-FEC limit (i.e., $3.8\times10^{-3}$) when the ROP is greater than about $-5.8$dBm using the uniform QAM, while it can achieve to be below 7\% HD-FEC limit when ROP is greater than approximately $-7.2$dBm using PCS QAM. After 50km SSMF transmission, not only the limited bandwidth, but also CD results in signal power fading, which leads to the BER performance degradation compared to the OBTB transmission scenario. For the 50km SSMF transmission scenario, when the ROP is greater than approximately $-4.8$dBm for the uniform QAM, BER can only achieve to be below the 20\% soft-decision FEC (SD-FEC) limit (i.e., $2.2\times10^{-2}$). If PCS QAM is utilized, BER is below the 20\% SD-FEC limit when ROP is greater than about $-6.8$dBm. Additionally, BER of the 100Gbit/s IM/DD optical multi-rate Nyquist-SCM system using PCS QAM scheme is below 7\% HD-FEC limit with ROP great than approximately $-4.5$dBm. The actual systems would likely employ the optical amplifiers, which are OSNR-limited scenarios. BER performance versus OSNRs is shown in Fig. \ref{BER}(b). The ROP is fixed at about $-4$dBm and OSNR is adjusted by varying the ASE noise power. The BER decreases with the increase of OSNR, which shows the similar trend to that with the increase of ROP in the ROP-limited scenarios.

\begin{figure}[!t]
	\centering
	\includegraphics[width = 2.95 in]{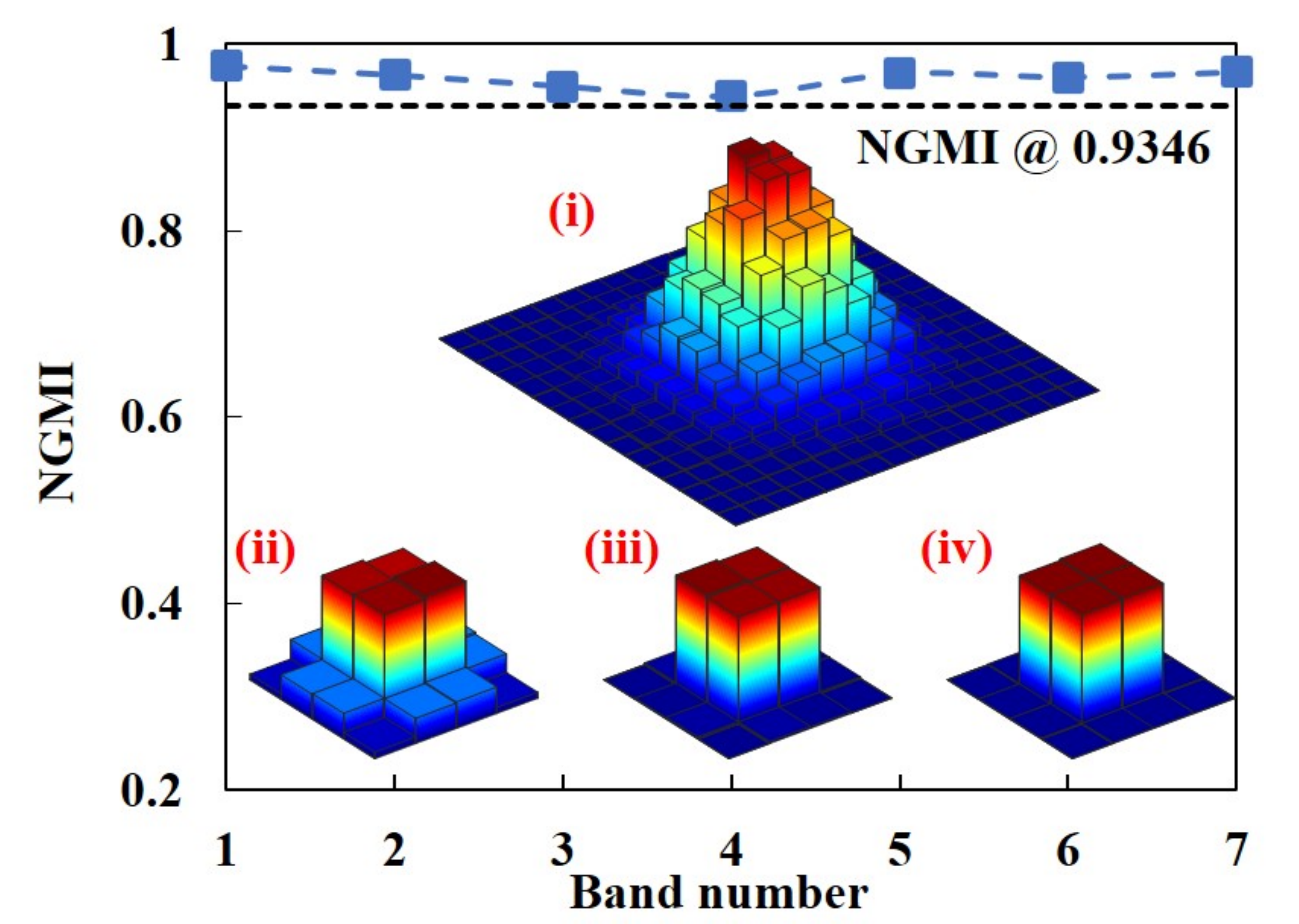}
	\caption{NGMI of 100Gbit/s IM/DD optical multi-rate Nyquist-SCM each band signal adopting PCS QAM over 50km SSMF transmission at ROP of approximately $-4$dBm. Insets are the probability of the constellation points of the (i) $1^{\rm{st}}$, (ii) $3^{\rm{rd}}$, (iii) $5^{\rm{th}}$ and (iv) $7^{\rm{th}}$ band, respectively.}
	\label{NGMI}
\end{figure}

To predict post-FEC BER performance, NGMI is more commonly used than the pre-FEC BER for the systems adopting PCS QAM, which can be estimated as \cite{cho2017normalized, chen2019adaptive}
\begin{equation}
\text{NGMI} \approx 1-\frac{1}{mS} \sum_{k=1}^{S} \sum_{i=1}^{m} \log _{2}\left(1+e^{(-1)^{b_{k, i}} \Lambda_{k, i}}\right)
\end{equation}
where $m = \log_2(M)$ and $S$ is the number of symbols. $b_{k, i}$ is the transmitted bits and $\Lambda_{k, i}$ represents the log-likelihood ratio. FEC overhead (OH) can be predicted by \cite{che2019squeezing}
\begin{equation}
\text{OH} = (1 - \text{NGMI}) / \text{NGMI}.
\end{equation}
The corresponding theoretical NGMI threshold of 7\% FEC OH is about 0.9346. Fig. \ref{NGMI} shows the NGMI of each band of 100Gbit/s IM/DD optical multi-rate Nyquist-SCM system adopting PCS QAM over 50km SSMF transmission at ROP of approximately $-4$dBm. NGMIs of each band are larger than 0.9346, which implies that the 7\% HD-FEC OH is enough for decoding. Insets show the probability of the constellation points of the (i) $1^{\rm{st}}$, (ii) $3^{\rm{rd}}$, (iii) $5^{\rm{th}}$  and (iv) $7^{\rm{th}}$ band, respectively. They are similar to those shown in Fig. \ref{Probability}.

\begin{figure}[!t]
	\centering
	\includegraphics[width = 2.95 in]{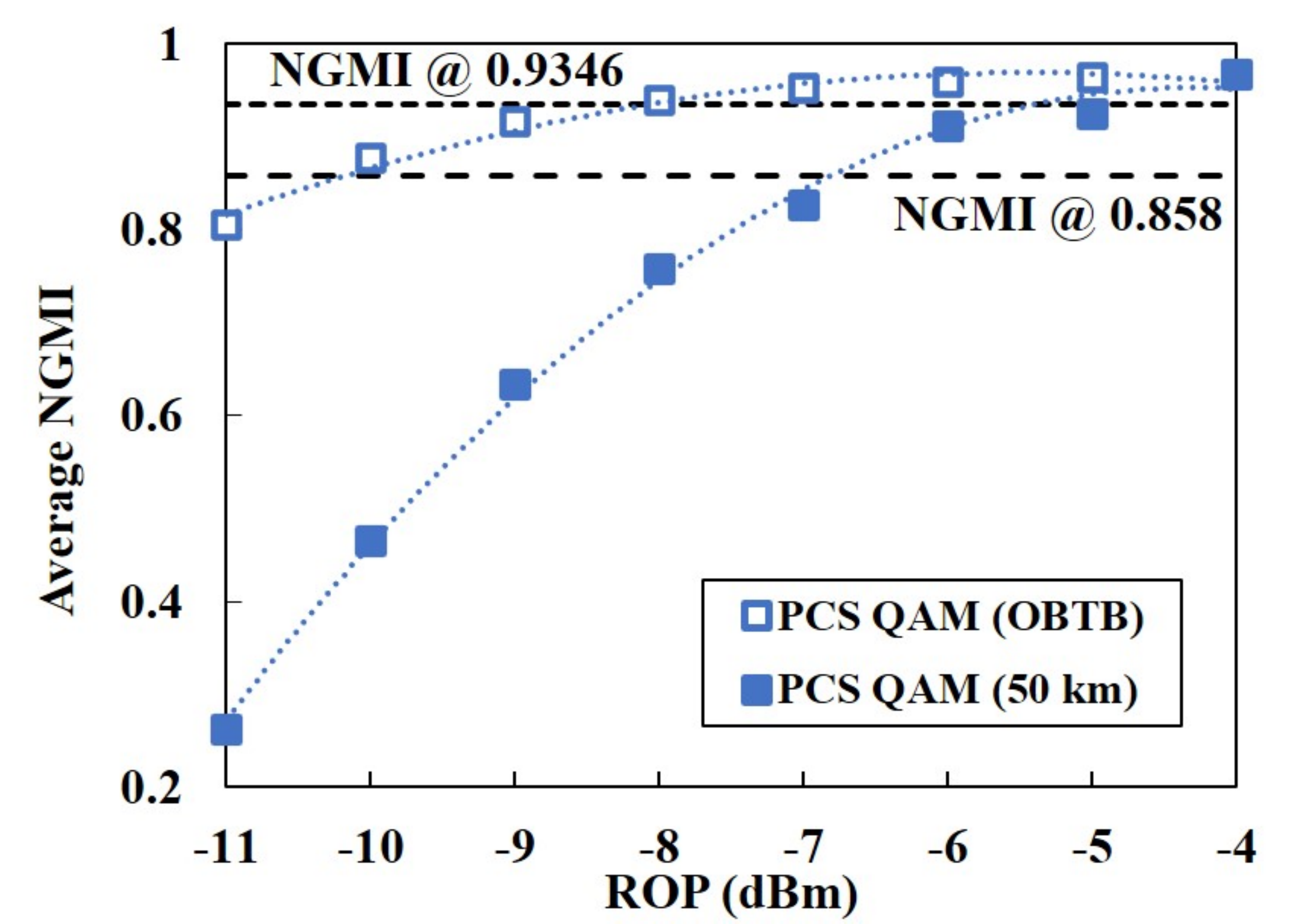}
	\caption{Average NGMI of the 100Gbit/s IM/DD optical multi-rate Nyquist-SCM system adopting PCS QAM over OBTB and 50km SSMF transmission.}
	\label{aveNGMI}
\end{figure}

Average NGMI is important for the multi-carriers system for the reason that channel coding and decoding are performed to the data on all subcarriers together. Average NGMI of such multi-rate SCM system can be calculated as \cite{chen2019single}
\begin{equation}
\text{NGMI}_{\text{avg}}=\frac{\sum_{i=1}^{N} RS_{i} \times m_{i} \times \text{NGMI}_{i}}{\sum_{i=1}^{N} RS_{i} \times m_{i}}.
\end{equation}
When 20\% low density parity check with 6.25\% staircase code is considered as the FEC code, the NGMI should be larger than 0.858 to achieve the $10^{-15}$ post-FEC BER \cite{alvarado2015replacing}. Fig. \ref{aveNGMI} shows the average NGMI versus ROPs of 100Gbit/s IM/DD optical multi-rate Nyquist-SCM system using PCS QAM over OBTB and 50km SSMF transmission. For OBTB transmission scenario, when ROP are equal to or greater than $-10$dBm and $-7$dBm, the average NGMIs are greater than the NGMI thresholds of 0.858 and 0.9346, respectively. For 50km SSMF transmission scenario, when ROP are equal to or greater than $-6$dBm and $-4$dBm, the average NGMIs are greater than the NGMI thresholds of 0.858 and 0.9346, respectively. When ROP is approximately $-4$dBm, the average NGMI is 0.967. Therefore, both pre-FEC BER and average NGMI indicate that the multi-rate Nyquist-SCM adopting PCS QAM scheme not only has a strong resistance to CD, but also makes full utilization of SNR to maximize the capacity-reach.

\section{Conclusion}
\label{Conclusion}
In this paper, to the best of our knowledge, we propose the first multi-rate Nyquist-SCM for C-band 100Gbit/s signal transmission over 50km dispersion-uncompensated link. Based on the characteristics of dispersive channel, an optical DSB Nyquist-SCM with multi-rate subcarriers is proposed to keep away from the CD-caused spectral nulls flexibly. As a result, the multi-rate Nyquist-SCM signal has a strong resistance to CD. Signal on each subcarrier can be recovered individually by ACMD algorithm, including the FFE with no more than 31 taps, a two-tap PF and MLSE with one memory length. Combining with entropy loading based on PCS to maximize the capacity-reach, the C-band 100Gbit/s multi-rate Nyquist-SCM signal over 50km dispersion-uncompensated link achieves 7\% HD-FEC limit and average NGMI of 0.967. In conclusion, the multi-rate Nyquist-SCM shows great potentials in solving CD-caused spectral distortions, which successfully combines the advanced technologies of Nyquist-SCM, multi-rate subcarriers, ACMD algorithm and entropy loading based on PCS.

\ifCLASSOPTIONcaptionsoff
  \newpage
\fi



\bibliographystyle{IEEEtran}
\bibliography{sample}

\begin{thebibliography}{10}
\providecommand{\url}[1]{#1}
\csname url@samestyle\endcsname
\providecommand{\newblock}{\relax}
\providecommand{\bibinfo}[2]{#2}
\providecommand{\BIBentrySTDinterwordspacing}{\spaceskip=0pt\relax}
\providecommand{\BIBentryALTinterwordstretchfactor}{4}
\providecommand{\BIBentryALTinterwordspacing}{\spaceskip=\fontdimen2\font plus
\BIBentryALTinterwordstretchfactor\fontdimen3\font minus
  \fontdimen4\font\relax}
\providecommand{\BIBforeignlanguage}[2]{{%
\expandafter\ifx\csname l@#1\endcsname\relax
\typeout{** WARNING: IEEEtran.bst: No hyphenation pattern has been}%
\typeout{** loaded for the language `#1'. Using the pattern for}%
\typeout{** the default language instead.}%
\else
\language=\csname l@#1\endcsname
\fi
#2}}
\providecommand{\BIBdecl}{\relax}
\BIBdecl

\bibitem{zhou2019beyond}
X.~Zhou, R.~Urata, and H.~Liu, ``{Beyond 1Tb/s datacenter interconnect
  technology: challenges and solutions},'' in \emph{Optical Fiber
  Communications Conference and Exhibition (OFC)}.\hskip 1em plus 0.5em minus
  0.4em\relax Optical Society of America, 2019, p. Tu2F.5.

\bibitem{zhong2018digital}
K.~Zhong, X.~Zhou, J.~Huo, C.~Yu, C.~Lu, and A.~P.~T. Lau, ``{Digital signal
  processing for short-reach optical communications: A review of current
  technologies and future trends},'' \emph{Journal of Lightwave Technology},
  vol.~36, no.~2, pp. 377--400, 2018.

\bibitem{cheng2018recent}
Q.~Cheng, M.~Bahadori, M.~Glick, S.~Rumley, and K.~Bergman, ``{Recent advances
  in optical technologies for data centers: a review},'' \emph{Optica}, vol.~5,
  no.~11, pp. 1354--1370, 2018.

\bibitem{liu2018dd}
G.~N. Liu, L.~Zhang, T.~Zuo, and Q.~Zhang, ``{IM/DD transmission techniques for
  emerging 5G fronthaul, DCI, and metro applications},'' \emph{Journal of
  Lightwave Technology}, vol.~36, no.~2, pp. 560--567, 2018.

\bibitem{chagnon2019direct}
M.~Chagnon, ``{Direct-detection technologies for intra-and inter-data center
  optical links},'' in \emph{2019 Optical Fiber Communications Conference and
  Exhibition (OFC)}.\hskip 1em plus 0.5em minus 0.4em\relax Optical Society of
  America, 2019, p. W1F.4.

\bibitem{wei2018challenges}
J.~Wei, Q.~Zhang, L.~Zhang, N.~Stojanovic, C.~Prodaniuc, F.~Karinou, and
  C.~Xie, ``{Challenges and advances of direct detection systems for DCI and
  metro networks},'' in \emph{2018 Optical Fiber Communications Conference and
  Exposition (OFC)}.\hskip 1em plus 0.5em minus 0.4em\relax Optical Society of
  America, 2018, p. W2A.60.

\bibitem{hong2021numerical}
Y.~Hong, K.~R. Bottrill, N.~Taengnoi, N.~K. Thipparapu, Y.~Wang, J.~K. Sahu,
  D.~J. Richardson, and P.~Petropoulos, ``{Numerical and experimental study on
  the impact of chromatic dispersion on O-band direct-detection
  transmission},'' \emph{Applied Optics}, vol.~60, no.~15, pp. 4383--4390,
  2021.

\bibitem{eiselt2016evaluation}
N.~Eiselt, J.~Wei, H.~Griesser, A.~Dochhan, M.~H. Eiselt, J.-P. Elbers,
  J.~J.~V. Olmos, and I.~T. Monroy, ``{Evaluation of real-time 8$\times$56.25
  Gb/s (400G) PAM-4 for inter-data center application over 80 km of SSMF at
  1550 nm},'' \emph{Journal of Lightwave Technology}, vol.~35, no.~4, pp.
  955--962, 2016.

\bibitem{zhang2017eml}
J.~Zhang, J.~Yu, and H.-C. Chien, ``{EML-based IM/DD 400G
  (4$\times$112.5-Gbit/s) PAM-4 over 80 km SSMF based on linear
  pre-equalization and nonlinear LUT pre-distortion for inter-DCI
  applications},'' in \emph{2017 Optical Fiber Communications Conference and
  Exhibition (OFC)}.\hskip 1em plus 0.5em minus 0.4em\relax Optical Society of
  America, 2017, p. W4I.4.

\bibitem{lee2016112}
J.~Lee, N.~Kaneda, and Y.-K. Chen, ``{112-Gbit/s intensity-modulated
  direct-detect vestigial-sideband PAM4 transmission over an 80-km SSMF
  link},'' in \emph{ECOC 2016; 42nd European Conference on Optical
  Communication}.\hskip 1em plus 0.5em minus 0.4em\relax VDE, 2016, p. M.2.D.3.

\bibitem{zhang2014c}
Q.~Zhang, Y.~Fang, E.~Zhou, T.~Zuo, L.~Zhang, G.~N. Liu, and X.~Xu, ``{C-band
  56Gbps transmission over 80-km single mode fiber without chromatic dispersion
  compensation by using intensity-modulation direct-detection},'' in \emph{2014
  The European Conference on Optical Communication (ECOC)}.\hskip 1em plus
  0.5em minus 0.4em\relax IEEE, 2014, p. P.5.19.

\bibitem{zhang2015c}
L.~Zhang, Q.~Zhang, T.~Zuo, E.~Zhou, G.~N. Liu, and X.~Xu, ``{C-band single
  wavelength 100-Gb/s IM-DD transmission over 80-km SMF without CD compensation
  using SSB-DMT},'' in \emph{2015 Optical Fiber Communications Conference and
  Exhibition (OFC)}.\hskip 1em plus 0.5em minus 0.4em\relax IEEE, 2015, p.
  Th4A.2.

\bibitem{wan201764}
Z.~Wan, J.~Li, L.~Shu, S.~Fu, Y.~Fan, F.~Yin, Y.~Zhou, Y.~Dai, and K.~Xu,
  ``{64-Gb/s SSB-PAM4 transmission over 120-km dispersion-uncompensated SSMF
  with blind nonlinear equalization, adaptive noise-whitening postfilter and
  MLSD},'' \emph{Journal of Lightwave Technology}, vol.~35, no.~23, pp.
  5193--5200, 2017.

\bibitem{li2018spectrally}
Z.~Li, M.~S. Erk{\i}l{\i}n{\c{c}}, K.~Shi, E.~Sillekens, L.~Galdino, T.~Xu,
  B.~C. Thomsen, P.~Bayvel, and R.~I. Killey, ``{Spectrally efficient 168
  Gb/s/$\lambda$ WDM 64-QAM single-sideband Nyquist-subcarrier modulation with
  Kramers--Kronig direct-detection receivers},'' \emph{Journal of Lightwave
  Technology}, vol.~36, no.~6, pp. 1340--1346, 2018.

\bibitem{le2019experimental}
S.~T. Le and K.~Schuh, ``{Experimental verification of equalization enhanced
  phase noise in Kramers-Kronig transmissions},'' in \emph{2019 Optical Fiber
  Communications Conference and Exhibition (OFC)}.\hskip 1em plus 0.5em minus
  0.4em\relax IEEE, 2019, p. Tu2B.2.

\bibitem{le20195}
S.~T. Le, K.~Schuh, R.~Dischler, F.~Buchali, L.~Schmalen, and H.~Buelow,
  ``{5$\times$510 Gbps single-polarization direct-detection WDM transmission
  over 80 km of SSMF},'' in \emph{Optical Fiber Communication
  Conference}.\hskip 1em plus 0.5em minus 0.4em\relax Optical Society of
  America, 2019, pp. Tu2B--1.

\bibitem{mecozzi2016kramers}
A.~Mecozzi, C.~Antonelli, and M.~Shtaif, ``{Kramers--Kronig coherent
  receiver},'' \emph{Optica}, vol.~3, no.~11, pp. 1220--1227, 2016.

\bibitem{zhang2016transmission}
Q.~Zhang, N.~Stojanovic, C.~Xie, C.~Prodaniuc, and P.~Laskowski,
  ``{Transmission of single lane 128 Gbit/s PAM-4 signals over an 80 km SSMF
  link, enabled by DDMZM aided dispersion pre-compensation},'' \emph{Optics
  Express}, vol.~24, no.~21, pp. 24\,580--24\,591, 2016.

\bibitem{erkilincc2015spectrally}
M.~S. Erk{\i}l{\i}n{\c{c}}, Z.~Li, S.~Pachnicke, H.~Griesser, B.~C. Thomsen,
  P.~Bayvel, and R.~I. Killey, ``{Spectrally efficient WDM Nyquist pulse-shaped
  16-QAM subcarrier modulation transmission with direct detection},''
  \emph{Journal of Lightwave Technology}, vol.~33, no.~15, pp. 3147--3155,
  2015.

\bibitem{torres2020100+}
P.~Torres-Ferrera, G.~Rizzelli, V.~Ferrero, and R.~Gaudino, ``{100+
  Gbps/$\lambda$ 50 km C-band downstream PON using CD digital pre-compensation
  and direct-detection ONU receiver},'' \emph{Journal of Lightwave Technology},
  vol.~38, no.~24, pp. 6807--6816, 2020.

\bibitem{tang2019c}
X.~Tang, S.~Liu, Z.~Sun, H.~Cui, X.~Xu, J.~Qi, M.~Guo, Y.~Lu, and Y.~Qiao,
  ``{C-band 56-Gb/s PAM4 transmission over 80-km SSMF with electrical
  equalization at receiver},'' \emph{Optics Express}, vol.~27, no.~18, pp.
  25\,708--25\,717, 2019.

\bibitem{xin202050}
H.~Xin, K.~Zhang, L.~Li, H.~He, and W.~Hu, ``{50 Gbps PAM-4 Over Up to 80-km
  Transmission With C-Band DML Enabled by Post-Equalizer},'' \emph{IEEE
  Photonics Technology Letters}, vol.~32, no.~11, pp. 643--646, 2020.

\bibitem{rath2017tomlinson}
R.~Rath, D.~Clausen, S.~Ohlendorf, S.~Pachnicke, and W.~Rosenkranz,
  ``{Tomlinson--Harashima precoding for dispersion uncompensated PAM-4
  transmission with direct-detection},'' \emph{Journal of Lightwave
  Technology}, vol.~35, no.~18, pp. 3909--3917, 2017.

\bibitem{zhou2020c}
J.~Zhou, H.~Wang, L.~Liu, C.~Yu, Y.~Feng, S.~Gao, W.~Liu, and Z.~Li, ``{C-band
  56 Gbit/s on/off keying system over a 100 km dispersion-uncompensated link
  using only receiver-side digital signal processing},'' \emph{Optics Letters},
  vol.~45, no.~3, pp. 758--761, 2020.

\bibitem{zhou2021processing}
J.~Zhou, H.~Wang, Y.~Feng, W.~Liu, S.~Gao, C.~Yu, and Z.~Li, ``{Processing for
  dispersive intensity-modulation and direct-detection fiber-optic
  communications},'' \emph{Optics Letters}, vol.~46, no.~1, pp. 138--141, 2021.

\bibitem{nadal2014dmt}
L.~Nadal, M.~S. Moreolo, J.~M. F{\`a}brega, A.~Dochhan, H.~Grie{\ss}er,
  M.~Eiselt, and J.-P. Elbers, ``{DMT modulation with adaptive loading for high
  bit rate transmission over directly detected optical channels},''
  \emph{Journal of Lightwave Technology}, vol.~32, no.~21, pp. 3541--3551,
  2014.

\bibitem{le2020100gbps}
S.~T. Le, T.~Drenski, A.~Hills, M.~King, K.~Kim, Y.~Matsui, and T.~Sizer,
  ``{100Gbps DMT ASIC for Hybrid LTE-5G Mobile Fronthaul Networks},''
  \emph{Journal of Lightwave Technology}, vol.~39, no.~3, pp. 801--812, 2021.

\bibitem{park2000papr}
M.~Park, H.~Jun, J.~Cho, N.~Cho, D.~Hong, and C.~Kang, ``{PAPR reduction in
  OFDM transmission using Hadamard transform},'' in \emph{2000 IEEE
  International Conference on Communications. ICC 2000. Global Convergence
  Through Communications. Conference Record}, vol.~1.\hskip 1em plus 0.5em
  minus 0.4em\relax IEEE, 2000, pp. 430--433.

\bibitem{cartledge2014100}
J.~C. Cartledge and A.~S. Karar, ``{100 Gb/s intensity modulation and direct
  detection},'' \emph{Journal of lightwave technology}, vol.~32, no.~16, pp.
  2809--2814, 2014.

\bibitem{erkilincc2016spectrally}
M.~S. Erk{\i}l{\i}n{\c{c}}, M.~P. Thakur, S.~Pachnicke, H.~Griesser,
  J.~Mitchell, B.~C. Thomsen, P.~Bayvel, and R.~I. Killey, ``{Spectrally
  efficient WDM Nyquist pulse-shaped subcarrier modulation using a dual-drive
  Mach--Zehnder Modulator and direct detection},'' \emph{Journal of Lightwave
  Technology}, vol.~34, no.~4, pp. 1158--1165, 2016.

\bibitem{erkilinc2015performance}
M.~S. Erk{\i}l{\i}nc, S.~Pachnicke, H.~Griesser, B.~C. Thomsen, P.~Bayvel, and
  R.~I. Killey, ``{Performance comparison of single-sideband direct detection
  Nyquist-subcarrier modulation and OFDM},'' \emph{Journal of Lightwave
  Technology}, vol.~33, no.~10, pp. 2038--2046, 2015.

\bibitem{bo2018toward}
T.~Bo and H.~Kim, ``{Toward practical Kramers-Kronig receiver: resampling,
  performance, and implementation},'' \emph{Journal of Lightwave Technology},
  vol.~37, no.~2, pp. 461--469, 2018.

\bibitem{vassilieva2019enabling}
O.~Vassilieva, I.~Kim, and T.~Ikeuchi, ``{Enabling technologies for fiber
  nonlinearity mitigation in high capacity transmission systems},''
  \emph{Journal of Lightwave Technology}, vol.~37, no.~1, pp. 50--60, 2019.

\bibitem{qiu2014subcarrier}
M.~Qiu, Q.~Zhuge, X.~Xu, M.~Chagnon, M.~Morsy-Osman, and D.~V. Plant,
  ``{Subcarrier multiplexing using DACs for fiber nonlinearity mitigation in
  coherent optical communication systems},'' in \emph{Optical Fiber
  Communication Conference}.\hskip 1em plus 0.5em minus 0.4em\relax Optical
  Society of America, 2014, p. Tu3J.2.

\bibitem{sun2020800g}
H.~Sun, M.~Torbatian, M.~Karimi, R.~Maher, S.~Thomson, M.~Tehrani, Y.~Gao,
  A.~Kumpera, G.~Soliman, A.~Kakkar \emph{et~al.}, ``{800G DSP ASIC design
  using probabilistic shaping and digital sub-carrier multiplexing},''
  \emph{Journal of lightwave technology}, vol.~38, no.~17, pp. 4744--4756,
  2020.

\bibitem{wang2020adaptive}
H.~Wang, J.~Zhou, D.~Guo, Y.~Feng, W.~Liu, C.~Yu, and Z.~Li, ``{Adaptive
  Channel-Matched Detection for C-Band 64-Gbit/s Optical OOK System Over 100-km
  Dispersion-Uncompensated Link},'' \emph{Journal of Lightwave Technology},
  vol.~38, no.~18, pp. 5048--5055, 2020.

\bibitem{zhou2016transmission}
J.~Zhou, L.~Zhang, T.~Zuo, Q.~Zhang, S.~Zhang, E.~Zhou, and G.~N. Liu,
  ``{Transmission of 100-Gb/s DSB-DMT over 80-km SMF Using 10-G class TTA and
  Direct-Detection},'' in \emph{ECOC 2016; 42nd European Conference on Optical
  Communication}.\hskip 1em plus 0.5em minus 0.4em\relax VDE, 2016, p.
  Tu.3.F.1.

\bibitem{zhou2021burst}
J.~Zhou, C.~Yang, D.~Wang, Q.~Sui, H.~Wang, S.~Gao, Y.-H. Feng, W.~Liu, Y.~Yan,
  J.~Li \emph{et~al.}, ``{Burst-Error-Propagation Suppression for
  Decision-Feedback Equalizer in Field-Trial Submarine Fiber-Optic
  Communications},'' \emph{Journal of Lightwave Technology}, vol.~39, no.~14,
  pp. 4601--4606, 2021.

\bibitem{wei2017multi}
J.~Wei and E.~Giacoumidis, ``{Multi-band CAP for next-generation optical access
  networks using 10-G optics},'' \emph{Journal of Lightwave Technology},
  vol.~36, no.~2, pp. 551--559, 2017.

\bibitem{cho2017normalized}
J.~Cho, L.~Schmalen, and P.~J. Winzer, ``{Normalized generalized mutual
  information as a forward error correction threshold for probabilistically
  shaped QAM},'' in \emph{2017 European Conference on Optical Communication
  (ECOC)}.\hskip 1em plus 0.5em minus 0.4em\relax IEEE, 2017, p. M.2.D.2.

\bibitem{chen2019adaptive}
X.~Chen, Y.~Chen, M.~Tang, T.~Tong, S.~Fu, and D.~Liu, ``{Adaptive Uniform
  Entropy Loading for SSB-DMT Systems},'' \emph{Journal of Lightwave
  Technology}, vol.~37, no.~23, pp. 5961--5970, 2019.

\bibitem{che2019squeezing}
D.~Che and W.~Shieh, ``{Squeezing out the last few bits from band-limited
  channels with entropy loading},'' \emph{Optics Express}, vol.~27, no.~7, pp.
  9321--9329, 2019.

\bibitem{chen2019single}
X.~Chen, S.~Chandrasekhar, J.~Cho, and P.~Winzer, ``{Single-wavelength and
  single-photodiode entropy-loaded 554-Gb/s transmission over 22-km SMF},'' in
  \emph{Optical Fiber Communication Conference}.\hskip 1em plus 0.5em minus
  0.4em\relax Optical Society of America, 2019, p. Th4B.5.

\bibitem{alvarado2015replacing}
A.~Alvarado, E.~Agrell, D.~Lavery, R.~Maher, and P.~Bayvel, ``{Replacing the
  soft-decision FEC limit paradigm in the design of optical communication
  systems},'' \emph{Journal of Lightwave Technology}, vol.~33, no.~20, pp.
  4338--4352, 2015.

\end{thebibliography}
%

%


\end{document}